**Self-generation of colligative properties at hydrophilic surfaces**

Martin Chaplin

**London South Bank University**

The generally accepted view of osmotic pressure is that it is a colligative property, along with freezing point depression, boiling point elevation and vapour pressure lowering. These properties ideally depend on the concentration of dissolved solute molecules. Osmotic pressure, however, is also generated, without any solute, at hydrophilic surfaces. Here is presented a rationale and explanation for this phenomenon.

**Background**

It is, perhaps, unsurprising that ion exchange surfaces can generate very high osmotic pressures of over 100 MPa in water,[1] as they createe high surface concentrations of counter-ions. Poly-ionic nanoparticles, with high surface area, produce such a great osmotic pressure that they can be used in practical desalination processes.[2, 3] However, it has been experimentally verified that uncharged hydrophilic surfaces can do this also, without the presence of counter-ions or solutes.[4, 5] Much work on the effect of hydrophilic surfaces on the mesoscopic properties of the adjoining aqueous solutions has been done in the laboratory of Gerald Pollack,[6, 7] and the phenomena have been confirmed by many other independent workers.[8-10] In essence, it has been found that the interfacial water next to ionic charged or neutral uncharged hydrophilic surfaces expels solutes to the bulk of the solution that may be several hundred microns away. These exclusion zones (named as EZ-water) can be visualized when low-molecular weight dyes, proteins, micron-sized microspheres or other solutes are used. Also the EZ-water seems to possess other physical properties such as absorption at 270 nm,[11] greater density, greater viscosity and negative charge compared with the bulk water. There is no generally-accepted explanation for this osmotic pressure phenomenon or these properties of EZ-water. A new explanation covering all these phenomena is given below, involving the generation of osmotic pressure, that is very simple, easily understood and potentially very important in a number of related fields.

**Proposal for the generation of osmotic pressure at aqueous interfaces**

Wherever water is present in solution it may be considered as being either 'bound' or 'free', although there will be transitional water between these states. When considering the colligative properties, 'water' is considered bound to any solute when it has a lower entropy compared with pure liquid water. Such water may be considered part of the solute and not part of the dissolving 'free' water.[12] As pure liquid water consists of a mixture containing low-density water, made up of extensively hydrogen bonded structures, and higher density water,[13-17] consisting of much smaller less extensive clusters, the proportions of 'bound' or 'free' water in pure liquid water can vary with more strongly-bound larger clusters with behaviour approaching that of 'bound' water. In bulk liquid water, the relative concentrations of the two aqueous forms is of no consequence as all the water behaves the same throughout. If volumes of the solution contain different proportions of strongly and weakly hydrogen-bonded water molecules (or put even more simply that there is more extensive clustering present), then these different volumes will show a difference with respect to their water activity and chemical potential. Normally any such instantaneous differences in water activity and chemical potential between different volumes within the same mass of liquid would rapidly cause liquid movement from one to the other in order to equalize these states and so remove the chemical potential differences. However, where there are surfaces interacting with the liquid water, the concentration of the more extensive hydrogen-bonded clusters within the surface layer may differ from the bulk values with the surface interactions preventing the potential equalization between bulk and surface volumes. When this occurs, the surface water has a different water activity and chemical potential to the bulk, leading to differences in osmotic pressure, and other colligative properties. The change in the chemical potential ($\mu_w$) is $-\{RTLn(x_{ws}) - RTLn(x_{wb})\}$ (that is, a negative energy term is added to the chemical potential when $x_{ws} < x_{wb}$) where $x_{ws}$ is the mole fraction of the 'free' water ($0 < x_{ws} < 1$) in the surface layer and $x_{wb}$ is the mole fraction of the 'free' water ($0 < x_{wb} < 1$) in the bulk liquid.

At hydrophilic surfaces, interactions between the surface and neighbouring water molecules fix the localised hydrogen bonding and this, together with steric factors, increases the cluster extent and lifetime.[18] As the 'free' water reduces as compared with its bulk value when the formation of longer-lived and more extensive hydrogen bonded clusters increases,[18] so the osmotic pressure increases. This increase in osmotic pressure next to the surface will displace solutes from the surface towards the bulk until its effect is equalled by the osmotic pressure of the solution or the system reaches a steady state. As the first effect of this solute expulsion is naturally the formation of an increased concentration band as expelled solute mixes with the prior solute concentration, the extent of the expulsion will affect the whole of the unstirred layer (~1-100 μm). Where hydrophilic microparticles or nanoparticles are suspended, their surfaces will necessarily cause mutually repulsive osmotic pressure effects that may result in the ordering of the particles within small volumes of the liquid.[19-20] It should be noted that osmotic drive does not require a membrane to separate the two solutions[21] provided there are two phases.[22] Here the two phases consist of the unstirred and stirred layers. In this context, the affected aqueous layer behaves similarly to that described for exclusion zone (EZ) water by Pollack and this is put forward as a simple explanation of his experimental data.[7, 11, 23] It also shows similarities with the experiments on autothixotropy,[24-28] where the viscosity of unstirred solutions increase with time. The increase in density at the interface, as found in EZ-water, has been explained previously by the increase in clustering causing the water to behave as though it is at a lower temperature, which has also been used to explain the ease with which this surface layer freezes. The presence of 270 nm absorption in the interfacial water, as described for EZ-water,[11] may be attributed to the delocalization of electrons within the extended clustering as hydrogen bonding is known to involve electron delocalisation.[29] These electron delocalisations are stabilised by the addition of electrons but not by protonation, so causing the charge separation seen at these interfaces.[30]

Another effect of interfaces is the formation of evanescent waves due to the internal reflection of electromagnetic radiation. The standing electromagnetic wave produced will interact with water molecules to stabilise a standing wave of hydrogen bonded clusters that will increase the local concentration and extent of hydrogen bonded clusters so increasing the above osmotic effect, in agreement with the experimental data.[31-34]

It appears that a similar effect on solutes to the one described for water may occur in other polar solvents that can form hydrogen bonds.[35] thus reinforcing the likelihood that a mechanism is acting that does not depend on the specific properties of water, such as the here-described colligative thermodynamics.